\documentclass[journal]{IEEEtran}

\usepackage{color}
\usepackage{url}

\ifCLASSINFOpdf
  \usepackage[pdftex]{graphicx}
  \DeclareGraphicsExtensions{.pdf}
  \usepackage[pdftex]{hyperref} 

\else
  \usepackage[dvips]{graphicx}
  \DeclareGraphicsExtensions{.eps}
  
\fi

\usepackage{amsmath}
\usepackage{array}
\usepackage[all]{hypcap}

\newcommand{\ud}{\mathrm{d}}
\newcommand{\Eoper}[1]{\mathbf{A}\{#1\}}
\newcommand{\bracket}[1]{\left( #1 \right)}

\begin{document}
\title{Bicoherence analysis of nonstationary and nonlinear processes}
\author{$^*$Peter~Zsolt~Poloskei$^{1,2}$,
        Gergely~Papp$^1$,
        Gabor~Por$^2$,
        Laszlo~Horvath$^3$,
        and Gergo~I.~Pokol$^2$ \\
        \textit{
$^1$Max Planck Institute for Plasma Physics, D-85748 Garching, Germany \\
$^2$Institute of Nuclear Techniques BME, H-1117 Budapest, Hungary \\
$^3$York Plasma Institute, Department of Physics, University of York, Heslington, York, YO10 5DD UK\\
}}


\maketitle

\begin{abstract}
Bicoherence analysis is a well established method for identifying the quadratic nonlinearity of stationary processes. However, it is often applied without checking the basic assumptions of stationarity and convergence. The classic bicoherence, unfortunately, tends to give false positives -- high bicoherence values without actual nonlinear coupling of different frequency components -- for signals exhibiting rapidly changing amplitudes and limited length. The effect of false positive values can lead to misinterpretation of results, therefore a more prudent analysis is necessary in such cases. This paper analyses the properties of bispectrum and bicoherence in detail, generalizing these quantities to nonstationary processes. A step-by-step method is proposed to filter out false positives at a given confidence level for the case of nonstationary signals. We present a number of test cases, where the method is demonstrated on simple physics-based numerical systems. 
The approach and methodology introduced in the paper can be generalized to lower and higher order coherence calculations.
\end{abstract}


\section{Introduction}
The need to identify the existence (or lack) of second order nonlinear interactions in dynamical systems is a widespread problem in numerous disciplines. Examples can be found in population dynamics \cite{cazelles03population}, geophysics \cite{angeo06lowfreq}, oceanology \cite{xie10seawaves} or in plasma physics \cite{manz15gam}. The basic idea is to investigate the three-wave coupling in a signal, or more accurately, to characterize the fraction of the signal-energy of two waves that is quadratically phase coupled to a third wave at the sum frequency. The data analysis methods for stationary (on the scale of a single investigated time slice \cite{bendat2011random}) processes are well developed \cite{kim79bispectral,kim80bispectral}. However, more care is necessary when working with nonstationary processes and transient signals. There have been a number of previous approaches to extend the validity of the bicoherence method to rapidly changing systems. However, all of these required quasi-stationary behaviour on some short timescale. Good examples are the wavelet-bicoherence \cite{milligen95nonlin}, the short-time Fourier transform based bicoherence \cite{chandran2012stftbicoh,pokol2013continuous}, or the method of appying adaptive windowing to select an optimal window length with maximized local coherence \cite{ombao08nonstat}. All of these, however, require quasi-stationarity on a short time scale, which is not always the case in various real-life systems.

This paper introduces an approach to characterize nonlinearity of nonstationary systems based on the ``classical'' bicoherence technique \cite{kim79bispectral,kim80bispectral}, complemented by statistical analysis. The goal is to estimate the significance level of the measured bicoherence for each frequency-frequency point, in order to filter likely false positives (type I error), i.e. values of high bicoherence despite the lack of actual phase coupling. The method is based on the estimation of the {\em random bicoherence probability density function}. A similar Monte Carlo based method has been used earlier to estimate the significance level of wavelet spectra by Torrence et.al. \cite{torrence98practical}.

Section~\ref{sec:stat-bicoh} describes bicoherence calculation for stationary processes based on Kim et.al. \cite{kim79bispectral}, and exposes the problem with nonstationary systems. In section~\ref{sec:nonstat-bicoh} a model for understanding the statistical properties of bicoherence calculations is introduced, and we present a method to detect false positives by estimating the significance level of the measured bicoherence. A validation using simple model systems follows in Section~\ref{sec:model-system}. Finally, implications for lower and higher order coherence calculations are discussed in Section~\ref{sec:discussion}.

\section{Bicoherence for stationary processes}\label{sec:stat-bicoh}
Bicoherence calculation is a standard method to investigate second order (i.e.: quadratic) nonlinearities, which appear as phase coupling between different Fourier components of a signal. We define the coupling condition with the following \eqref{eq:phaseCouplingCondition1}-\eqref{eq:phaseCouplingCondition2} equations \cite{kim79bispectral}:
\begin{align}
	f_3 = f_1 + f_2, \label{eq:phaseCouplingCondition1}\\
	\varphi_3 = \varphi_1 + \varphi_2 & + \mathrm{const}, \label{eq:phaseCouplingCondition2}
\end{align}
where $f_i$ ($i \in \{ 1,2,3 \} $) is the frequency and $\varphi_i$ is the phase of the corresponding Fourier components of the signal. Quadratic nonlinear coupling occurs between  $f_1$ and $f_2$, generating a third component at the sum frequency $f_3$.
Throughout the paper we are going to work with Fourier components of real (i.e. not complex) signals, and note these Fourier components with capital letters. We approximate the analytic Fourier transformation with the following expression:
\begin{align}\label{eq:fft_basic}
	X(f) &= \lim\limits_{T \to \infty}\dfrac{1}{T}\int^{T/2}_{-T/2}x(t)e^{-j2\pi f t} \ud t = \nonumber\\
	 &= \lim\limits_{N \to \infty}\dfrac{1}{N}\sum_{i=1}^N \dfrac{1}{T}\int^{t_i+T/2}_{t_i-T/2}x(t)e^{-j2\pi f t} \ud t = \nonumber\\
	 &= \lim\limits_{N \to \infty}\dfrac{1}{N}\sum_{i=1}^N X^{(i)}(f) = \mathbf{E}\left(X^{(i)}(f)\right),
\end{align}
where $j$ is the imaginary unit vector, in the second line $t_i$ defines the locations of independent time windows. This definition assumes an infinite long $x(t)$ signal from which infinite number of $X^{(i)}$ independent realizations of the spectrum can be calculated. 
For infinite long stationary signals (originating from ergodic processes) the time average is identical to the average of the independent realizations. In reality, infinite long signals do not exist, therefore we approximate the transform with a ''sufficient" number of averages:
\begin{equation}\label{eq:fft}
	X(f) \approx \dfrac{1}{N}\sum_{i=1}^N X^{(i)}(f).
\end{equation}
This means that we divide the measured $x(t)$ signal into $N$ pieces of $T$ long time slices, and carry out Fourier transformation separately, before averaging the spectra. We will use ensemble average throughout this work. To emphasise this, we introduce the $\Eoper{.}$ averaging operator:
\begin{equation}
	\Eoper{F}=\dfrac{1}{N}\sum\limits_{i=1}^{N} F^{(i)}, \label{eq:Eoper}
\end{equation}
where $F^{(i)}$ marks a derived quantity (e.g.: $X^{(i)}$ Fourier-transform) from the $i$-th time slice. 
 
\begin{figure}[!hbt]
	\centering
		\includegraphics[width=0.85\linewidth]{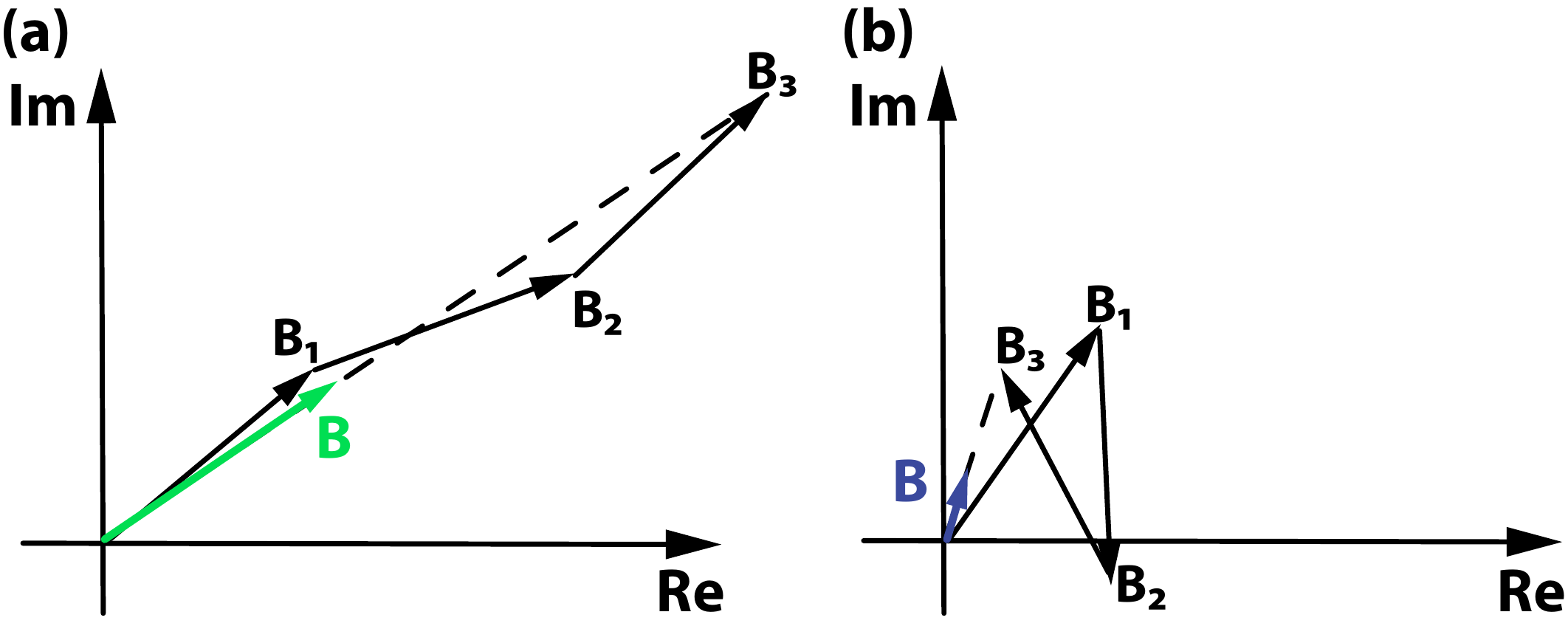}
		\caption{Bispectrum calculation visualized as a walk and averaging on the complex plane for given $(f_1,f_2)$ components. \textbf{(a)} If the two frequencies are phase coupled, their phase relation with the third generated component will remain constant over time (as described in eq. \eqref{eq:phaseCouplingCondition1}). Therefore the average members will align and the mean complex vector will have a large absolute value. \textbf{(b)} If the components are not phase coupled the phase of each consecutive step is random, and therefore the average value will be small.}
	\label{fig:walkOnComplexPlane}
\end{figure}

In the limit of ergodic stationary processes $\Eoper{.}$ will lead to the usual definition of the expected value, which we used in equation \eqref{eq:fft_basic}.
For the analysis of nonlinear quadratic coupling, first we define the {\em bispectrum} \cite{kim79bispectral}:
\begin{equation}\label{eq:bispectrumDef}
	B(f_1,f_2) = \Eoper{ X(f_1)X(f_2)X^*(f_1+f_2)},
\end{equation}
where $^*$ marks complex conjugation. Bispectrum is defined on a frequency-frequency plane. The procedure by which it is estimated can be visualized on the complex plane, illustrated in figure~\ref{fig:walkOnComplexPlane}.
In the case of phase coupling between components $(f_1,f_2)$, the absolute value of the bispectrum $B(f_1,f_2)$ will be large; otherwise its value will tend to zero with increasing number of averages.
{\em Bicoherence} is then defined as the bispectrum normalized in the following way:
\begin{equation}
	b^2(f_1,f_2) = \dfrac{|B(f_1,f_2)|^2}{\Eoper{| X(f_1)X(f_2) |^2} \Eoper{ | X(f_1+f_2) |^2 }}. \label{eq:bicoherenceDef}
\end{equation}
With the definition in eq.~\eqref{eq:bicoherenceDef},  $b(f_1,f_2) \in [0,1]$, and can be interpreted similarly to classic coherence \cite{kim79bispectral}.
If $(f_1,f_2)$ components are coupled (the conditions \eqref{eq:phaseCouplingCondition1}-\eqref{eq:phaseCouplingCondition2} are met), $b(f_1,f_2)\rightarrow 1$, else it will tend to zero as the number of averaging tends to infinity.

In the case of signals with finite length the frequency domain of the bicoherence is bounded with the Nyquist-frequency \cite{nyquist28certain}, which is shown as the red hexagonal frame in figure~\ref{fig:bicoherenceSymmetries}.
Due to the symmetries of the bispectrum (detailed in appendix \ref{sec:symmetries}) in the case of real signals it is enough to plot an even smaller region, which contains all the independent information (marked with ``{\bf P}'' in figure~\ref{fig:bicoherenceSymmetries}). For further analysis we will only present results in this area.
\begin{figure}[!hbt]
	\centering
		\includegraphics[width=0.6\linewidth]{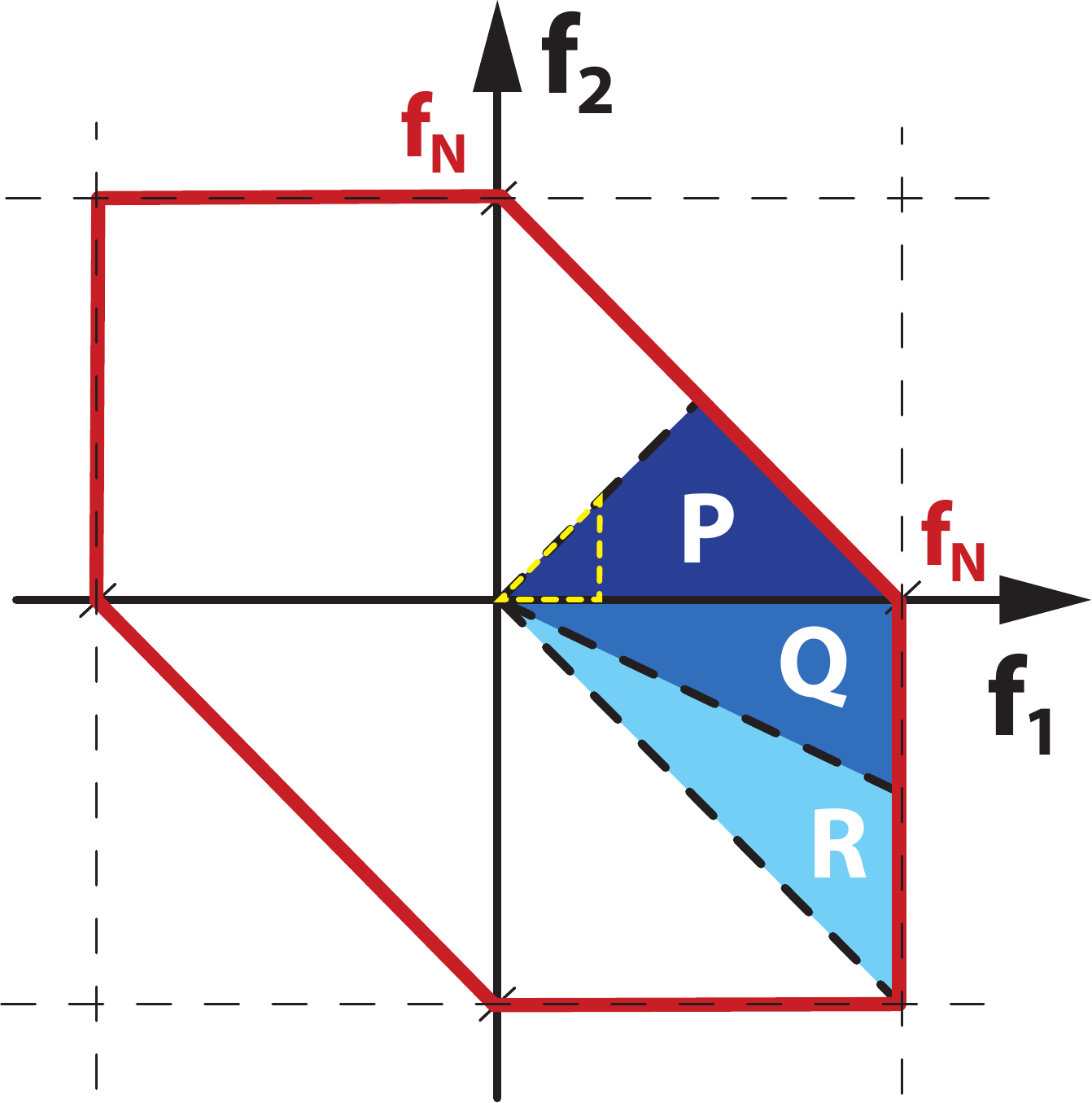}
		\caption{Bicoherence is defined on the $f_1,f_2$ plane. In case of finite signals it is bounded by the Nyquist-frequency, marked with red solid lines. Symmetries that arise from the definition reduce the plotting area to the ones marked with \textbf{P}, \textbf{Q}, \textbf{R}. Each of these contain equivalent information, therefore in the this paper we only plot the area marked with \textbf{P}. }
	\label{fig:bicoherenceSymmetries}
\end{figure}

\subsection{Numerical implementation}\label{sec:implementation}
Real-life digital signals sampled at a given $f_s$ sampling frequency have finite length (with $M$ discrete elements). The calculation of the averages in \eqref{eq:bispectrumDef} \& \eqref{eq:bicoherenceDef}, based on the definition in eq.~\eqref{eq:Eoper} requires the signal to be divided it into $N$ parts with equal $T=M/N/f_s$ length. The $T$ length of a single part and the $f_s$ sampling frequency determine the $f_N$ Nyquist frequency and $\Delta f$ frequency resolution of the calculations:
\begin{align}\label{eq:freqRes}
	f_N &= \dfrac{f_s}{2} \\	
	\Delta f &= f_s\dfrac{N}{M}.
\end{align}
It means that increasing the $N$ number of averages can only be achieved at the cost of degrading the frequency resolution, making a trade-off necessary for any given phenomenon under investigation. From a practical point of view, first we should consider the minimal  $\Delta f$ frequency resolution required for the investigated phenomena, which will then set the maximum $N$ number of averages.
Splitting the signal into parts (practically windowing with a boxcar window function) leads to sidelobes and other undesired features in the Discrete Fourier Transform. For the suppression of these effects it is beneficial to apply window functions to the smaller parts. Throughout the paper we utilize the {\em Hann window} with $50\%$ overlap \cite{welch67use}, where the window function is defined as
\begin{equation}\label{eq:hannWindow}
	w_\mathrm{Hann}(t) = \sin^2 \left( \dfrac{\pi}{T}t  \right),
\end{equation}
extending for a single period between two zero function values.

Let us now introduce the $\vec{X}^{(i)}$ Discreet Fourier Transformed (DFT) complex vector of the signal's $i$-th part, which has $2n=M/N$ elements, and its frequency resolution is defined by \eqref{eq:freqRes}. A single element of this vector will be marked as $X^{(i)}_k$, where $k \in (1, \cdots n)$ and the $k$-th element corresponds to the $k\cdot\Delta f$ frequency component. (Note that, as far as we are analysing real signals, we consider only the positive frequencies of the transform). Most programming languages support optimized matrix operations, therefore it is beneficial to rearrange our calculations in a matrix form to make calculations faster. First we define the $\mathbf{C}^{(i)}$ {\em cross-frequency matrix} with the following equation:
\begin{equation}\label{eq:crossfreq_matrix}
	\mathbf{C}^{(i)}_{kl} = \left(\vec{X}^{(i)} \otimes \vec{X}^{(i)}\right)_{kl} = X^{(i)}_kX^{(i)}_l,
\end{equation}
which is the dyadic product of $\vec{X}^{(i)}$ with itself. Then we define the $\mathbf{S}^{(i)}$ cyclically shifted frequency matrix of the $X^{(i)*}_j$ complex conjugate vector elements :
\begin{align}
\mathbf{S}^{(i)}_{kl} &=
	\begin{cases}
		X^{(i)*}_{k+l}, & \text{if}\ k \leq l\ \text{and}\ k+l \leq n  \\
		0, & \text{otherwise}
	\end{cases}
  \\
 &=
 \begin{pmatrix}
  X^{(i)*}_{1+1} & X^{(i)*}_{1+2} & \cdots & X^{(i)*}_{1+(n-2)} & X^{(i)*}_{1+(n-1)} & 0 \\
  0 & X^{(i)*}_{2+2} & \cdots & X^{(i)*}_{2+(n-2)} & 0  & 0\\
  \vdots  & \vdots  & \ddots & \vdots  & \vdots & \vdots \nonumber\\
  0 & 0 & \cdots & 0 & 0 & 0
 \end{pmatrix}
\end{align}
where the $k$-th row of the matrix is generated by shifting the $\vec{X}$ vector cyclically by $-k$, and finally zeroing out all elements outside of the upper quadrant defined by $k \leq l$ and $k+l \leq n$. Taking the element-wise product of the above defined matrices will produce the $\mathbf{B}$ bispectrum matrix, and these matrices can be used for the normalization to produce the $\mathbf{b}$ bicoherence. These statements are summarized in the following equations:
\begin{align}
\mathbf{B} &= \Eoper{\mathbf{C} \circ \mathbf{S}}\\
\mathbf{b}^2 &= \dfrac{|\mathbf{B}|^2}{\Eoper{|\mathbf{C}|^2}\circ\Eoper{|\mathbf{S}|^2}},
\end{align}
where $\circ$ denotes component-wise multiplication, and the $|.|$ absolute value, division and square of $\mathbf{C}$ and $\mathbf{S}$ are taken component-wise as well at the non-zero elements. Utilizing the matrix implementation on a modern computer can significantly speed up the calculation (the exact speedup depends on problem size, numerical library, and architecture).

\section{Bicoherence for nonstationary processes}\label{sec:nonstat-bicoh}
\begin{figure}[!hbt]
	\centering
		\includegraphics[width=0.85\linewidth]{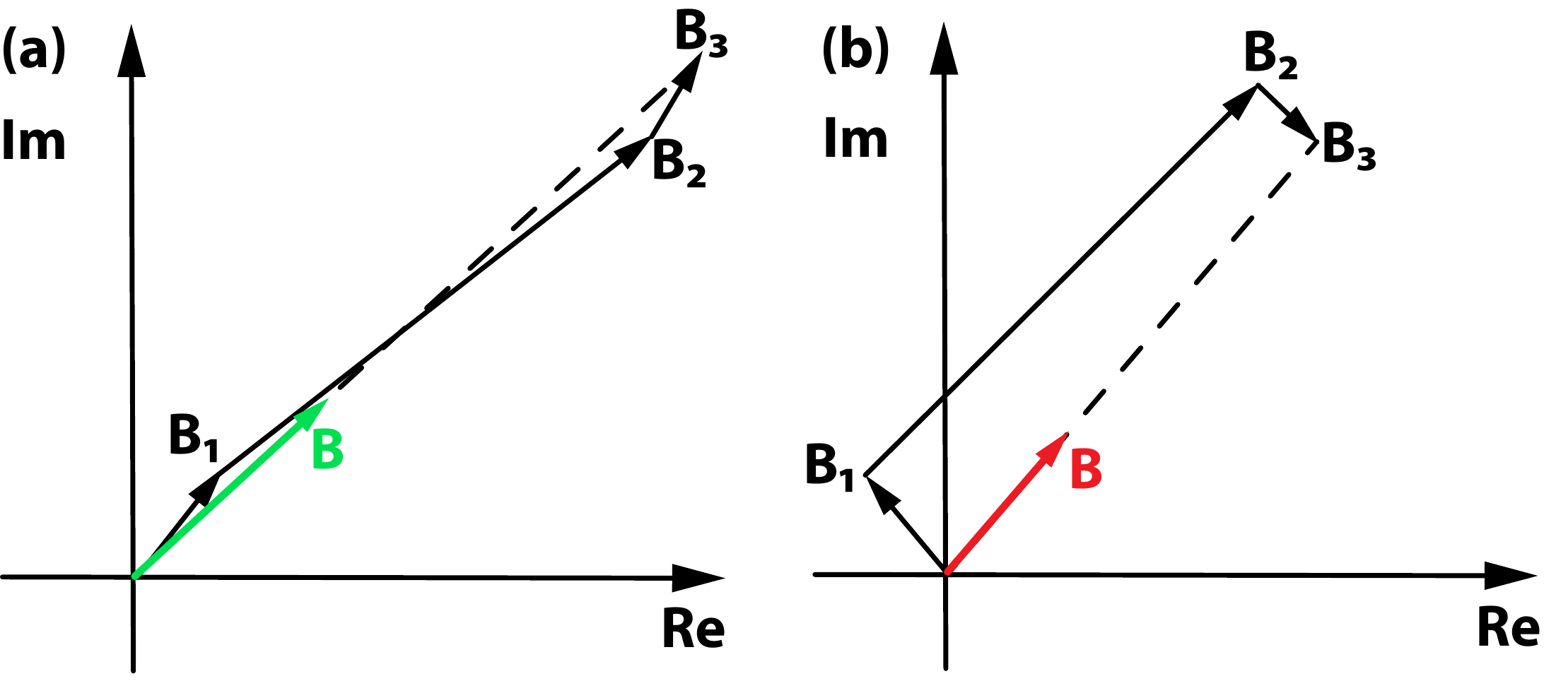}
		\caption{Bispectrum calculation for nonstationary signals as a walk on the complex plane for given $f_1$,$f_2$ components. \textbf{(a)} If the two frequencies are phase coupled, the average of the complex vectors will have a large absolute value. \textbf{(b)} If the phases are randomized but the amplitudes have a nonstationary characteristic the average value can still be high, leading to high bicoherence estimate (false positive).}
	\label{fig:walkOnComplexPlaneInstac}
\end{figure}

The calculations introduced so far were developed to investigate the non-linear properties of stationary processes \cite{kim79bispectral, kim80bispectral}. Problems in interpretation may arise when basic assumptions -- such as the requirement of stationary input signal -- are not fulfilled.
First, we demonstrate the characteristic problem of calculating bicoherence by definition \eqref{eq:bicoherenceDef} for nonstationary processes. In figure~\ref{fig:walkOnComplexPlaneInstac} we return to the visualization introduced in figure~\ref{fig:walkOnComplexPlane}, which illustrates the bispectrum calculation as averaging vectors on the complex plane. In the case of nonstationary processes, the length of the steps taken --which depend on the Fourier amplitudes -- may vary. In the case of phase coupling, the vectors align in the same direction. Therefore the bicoherence value is high, similarly as for stationary signals, as illustrated in figure \ref{fig:walkOnComplexPlaneInstac}(a). In the case of a nonstationary signal, the amplitudes in the blocks to be averaged vary significantly. This introduces a bias to the estimate, which may result in high bicoherence values even without phase coupling, as illustrated in \ref{fig:walkOnComplexPlaneInstac}(b). Therefore it is important to understand the effect of the distribution of Fourier amplitudes on the classical bicoherence calculation.

\subsection{Statistical properties of bicoherence}\label{sec:statistics}
Let us now analyse the statistical properties of bicoherence calculated for a real life (measurable) signal with arbitrary Fourier amplitude distribution. As a simple demonstration we model two processes -- one stationary with constant Fourier amplitude in time, the other nonstationary with time-varying amplitudes --, both of these without phase coupling. We then generate the windowed Fourier components (at a selected frequency) of the signals in the following way. For the stationary process the amplitude is constant in all time windows, while the nonstationary process has a uniform random distributed amplitude in the $[0,1]$ range. In each time window the phase of both processes is a uniform random variable $\varphi_i \in \left[0,2\pi\right]$. (This realization is analogous to the random walk picture illustrated in figures~\ref{fig:walkOnComplexPlane} \& \ref{fig:walkOnComplexPlaneInstac}).

\begin{figure}[!hbt]
	\centering
		\includegraphics[width=0.85\linewidth]{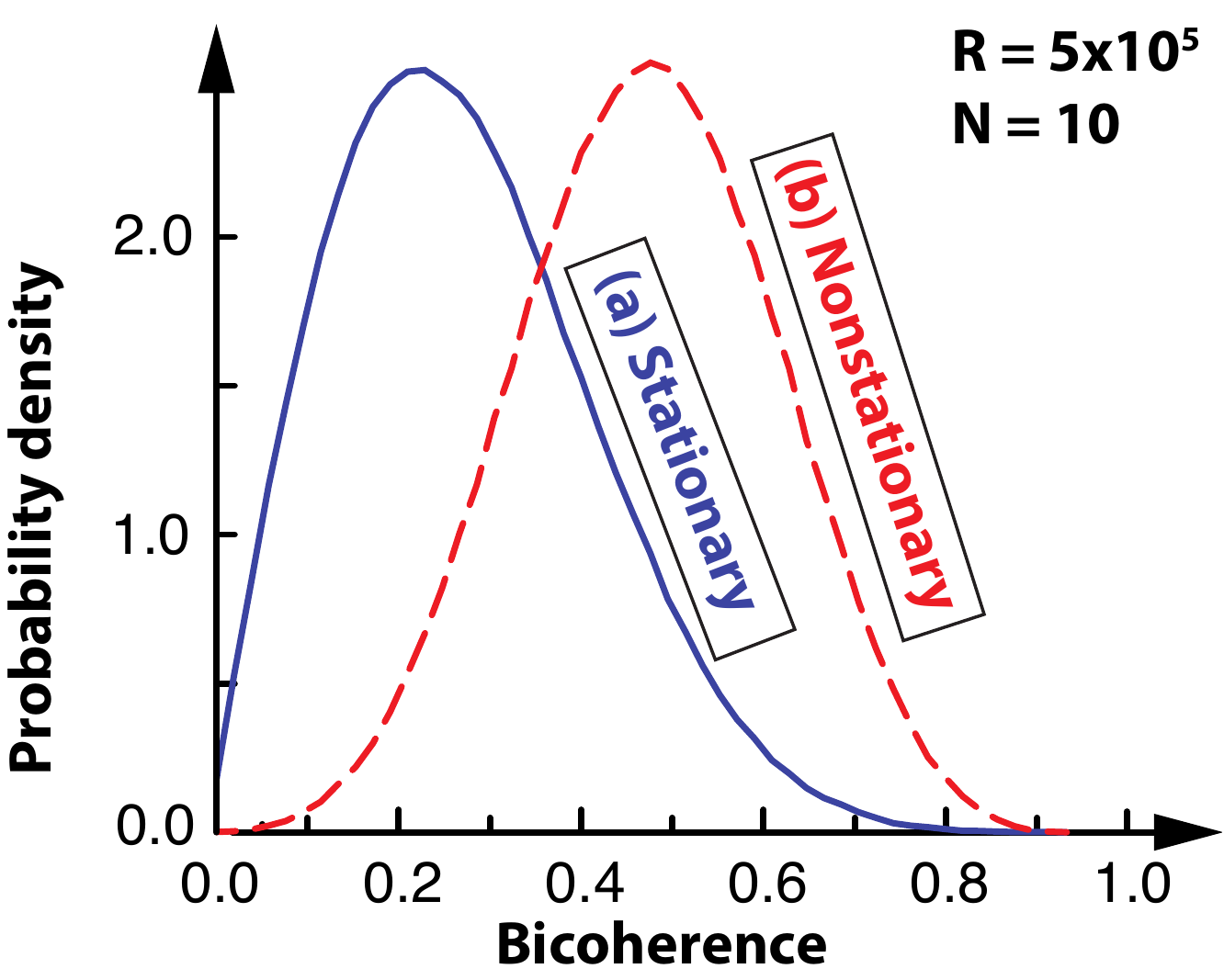}
		\caption{Probability density functions calculated for a stationary and a nonstationary process as illustrated in figures \ref{fig:walkOnComplexPlane}b and \ref{fig:walkOnComplexPlaneInstac}b. In the case of nonstationary process (marked with dashed line) the expected value of the bicoherence is significantly higher than in the case of the stationary process (solid line). This illustrates that in the case of nonstationary signals a high bicoherence value is more likely even without phase coupling.}
	\label{fig:randomPhaseDensity}
\end{figure}

In this particular example (shown in figure~\ref{fig:randomPhaseDensity}), for both processes $N=10$ blocks are generated for a single realization (to simulate short real-life signals) and then bicoherence is calculated using (\ref{eq:bicoherenceDef}), that corresponds to the specific realization of these random processes. Repeating this process a large $R \gg 1$ number of times (in this example: $R = 5{\cdot}10^5$) with newly generated phases represents further different realizations of the random processes. With this Monte Carlo method it is now possible to numerically estimate (using a histogram) the {\em random bicoherence probability density function}, hereinafter abbreviated as $\rho(b)$. A different $\rho(b)$ distribution corresponds to each $f_1,f_2$ frequency pair on the frequency-frequency plane (see e.g. figure~\ref{fig:bicoherenceSymmetries}). The $\rho(b)$ describes the bicoherence probability density of a random process with given Fourier amplitudes in the absence of phase coupling. For these two specific test signals described above, the estimated $\rho(b)$ is shown in figure~\ref{fig:randomPhaseDensity}. 
As we show in figure \ref{fig:randomPhaseDensity} -- for a fixed number of $N=10$ averages -- there is significant difference between the characteristics of the two different $\rho(b)$ distributions. For example, in the stationary case the expectation value is much lower than in the case of the nonstationary process. The generated $\rho(b)$ creates the basis for comparing the measured value of bicoherence with the bicoherence distribution of a process with similar amplitude, but completely random phase. In the following we will discuss, how the $\rho(b)$  probability density function can be used to identify possible false positives.

\subsection{Application of confidence filtering}\label{sec:confidence}

Numerically estimating $\rho(b)$ for a given amplitude distrubution gives us the possibility to analyse the bicoherence distribution in the absence of phase coupling. In practice, $\rho(b)$ will act as a reference to which the measured value can be compared to. With this in mind, the bicoherence analysis of the $f_1,f_2$ components of an arbitrary (even nonstationary) signal goes as the following:
\begin{enumerate}
\item We calculate the $X(f_i)$ Fourier amplitudes, for the given $f_1$, $f_2$ components.
\item Calculate the corresponding $b(f_1,f_2)$ measured bicoherence value, using definition (\ref{eq:bicoherenceDef}).
\item Generate the $\rho_{f_1,f_2}(b)$ phase-randomized bicoherence density function using the real life Fourier amplitudes of the the given frequency components with a random phase, using a sufficiently large (e.g.: $R = 2000$) number of realizations.
\item Define an $\alpha$ confidence level (e.g.: $\alpha=0.9$).
\item Calculate $b^c(\alpha)$ critical bicoherence value:
\begin{equation}
	\alpha = \int_0^{b^c} \rho(b)\ud b.
\end{equation}
\item If $b \ge b^c$, then it can be stated that at $\alpha$ confidence level $b$ does \textbf{NOT} originate from a random process.
\item If $b < b^c$, we assume that at $\alpha$ confidence level the bicoherence in that point is not significant.
\end{enumerate}

Naturally, we can carry out the above process for the entire frequency range of interest. The procedure to estimate $\rho_{f_1,f_2}(b)$ can be trivially paralellized for all $f_1,f_2$ couples.

\begin{figure}[!hbt]
	\centering
		\includegraphics[width=0.85\linewidth]{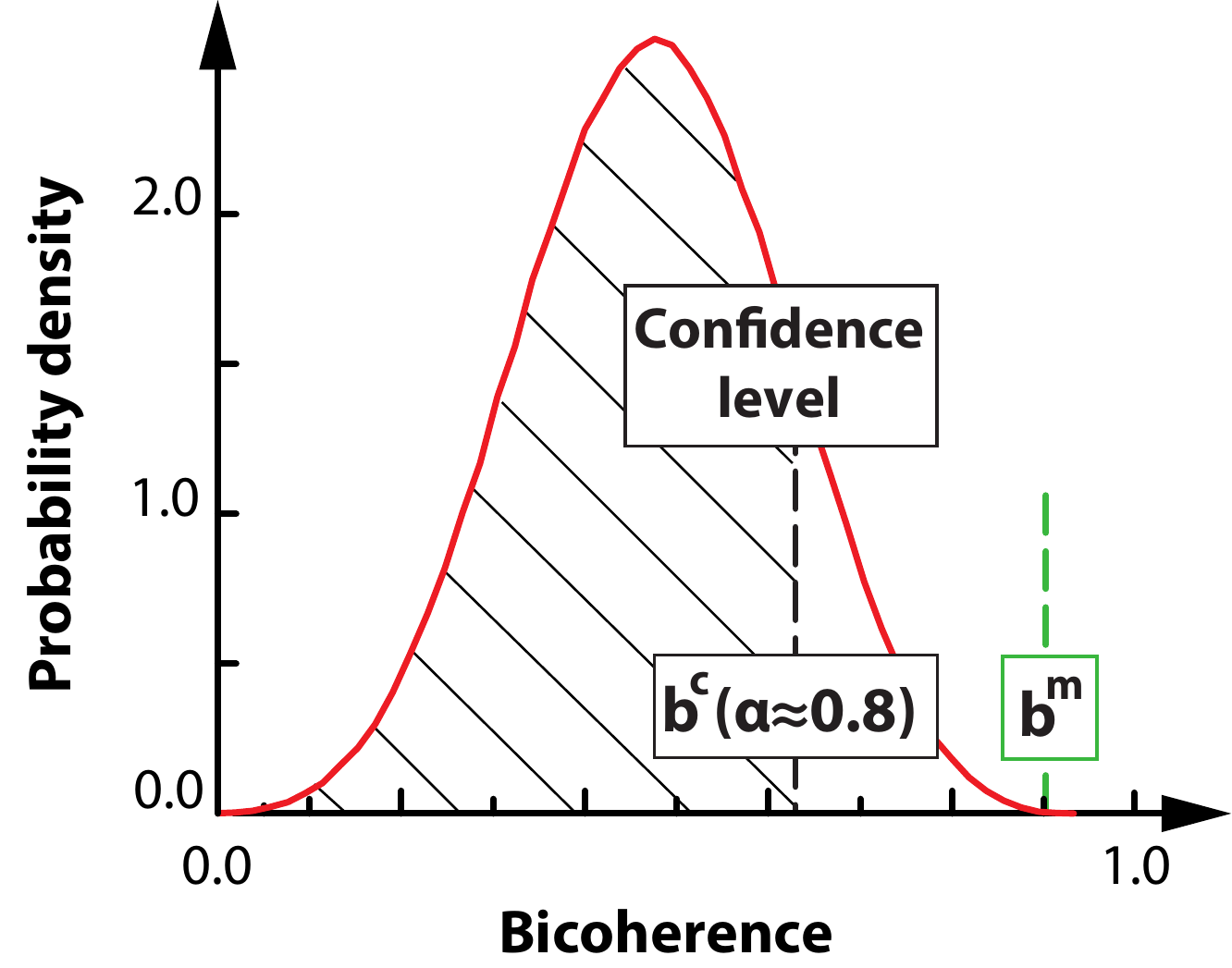}
		\caption{Visualization of the $\rho(b)$ phase-randomized bicoherence probability density function, the calculation of the $b^c$ critical value, and its comparison to the $b^m$ measured bicoherence.}
	\label{fig:confidenceLevel}
\end{figure}

Steps 3)--7) of the bicoherence analysis procedure are visualized in figure~\ref{fig:confidenceLevel}. We chose a global $\alpha$ confidence level, which defines a critical bicoherence value through the probability density function. In the case shown in figure~\ref{fig:confidenceLevel} the measured $b^m > b^c(\alpha)$ bicoherence value is higher than the critical one, therefore we \textbf{accept} it at the chosen confidence level as a sign (and measure) of phase coupling.

In section~\ref{sec:model-system} we will use different model systems to test the method introduced above, and to demonstrate the effect of different transients on the bicoherence calculation.

\section{Analysis of simple model systems}\label{sec:model-system}
\begin{figure}[!hbt]
	\centering
		\includegraphics[width=0.85\linewidth]{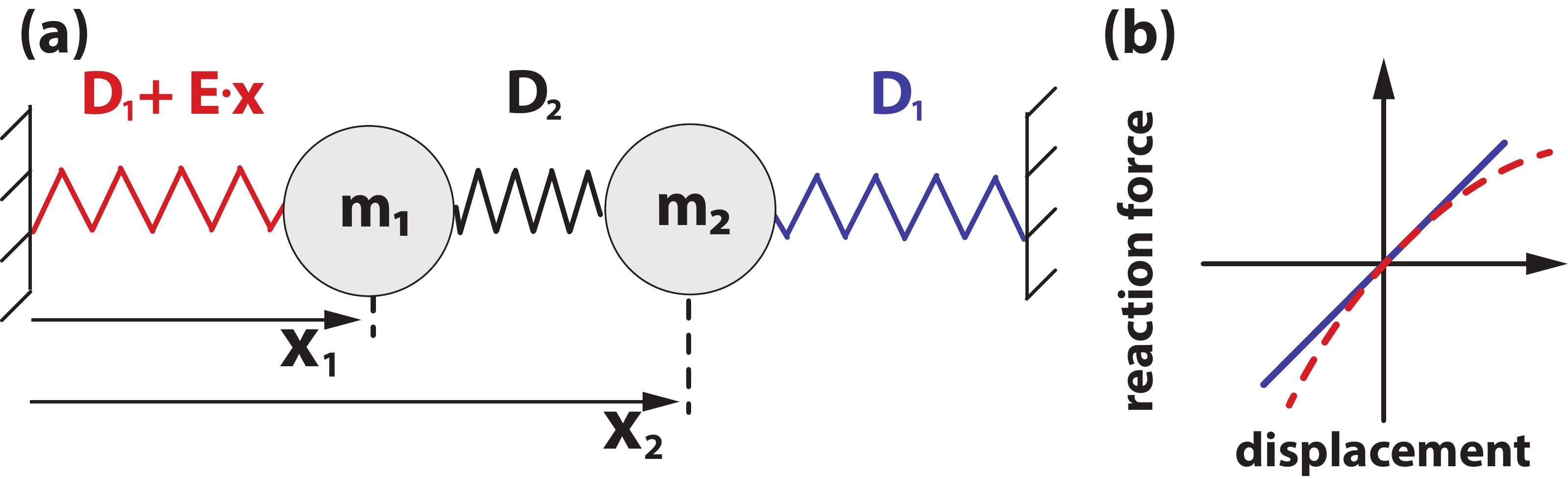}
		\caption{Sketch of the model system. Two $m_i$ masses attached to eachother and the wall with springs (with $D_i$ spring constants). One of the strings has a tunable $E$ nonlinearity parameter. The time signal used in the analysis is the $x_1(t)$ displacement of the first body.}
	\label{fig:modelSystem}
\end{figure}

We chose to simulate a simple physics model system (illustrated in figure~\ref{fig:modelSystem}) in order to validate the suggested method. The model consists of two masses $m_1$ and $m_2$ attached to each other and to the walls with springs, where two springs are ideal and one of the springs was chosen to have nonlinear characteristics described by the parameter $E$.

\noindent The equations of motion for the model system are:
\begin{align}
&m_1 \ddot{x}_1 = -D_1 x_1 + D_2 (x_2 - x_1) + E x_1^2 \nonumber \\ 
&m_2 \ddot{x}_2 = -D_1 x_2 - D_2 (x_2 - x_1), \label{eq:nonlinOscillation}
\end{align}
where $\ddot{x_i} = \ud ^2 x_i / \ud t^2$. Setting $E=0$ and $m_1=m_2=m$ we can calculate the eigenfrequencies of the investigated system:
\begin{align}
\omega_1^2 &=\dfrac{D_1}{m} \\
\omega_2^2 &=\dfrac{2D_2+D_1}{m}.
\end{align}
The time signal used in the analysis is the $x_1(t)$ displacement of the first body.
We can choose $\omega_1$ and $\omega_2$ as input parameters ($\omega_i = 2 \pi f_i$), and we can calculate the corresponding $D_1$ and $D_2$ constants, using e.g. $m = 1$ kg. For convenience we have parametrized the system such that the eigenfrequencies are far enough from eachother at $f_1= 45$~Hz and $f_2= 150 $~Hz. Nonlinearities are controlled by varying the $E$ parameter. Initial values for the calculations were $x_1(0)=1$, $\dot{x}_1(0)=0$, $x_2(0)=0$, $\dot{x}_2(0)=0$. The differential equation system was solved numerically with 4th order Runge-Kutta method, implemented in \texttt{IDL} language. $15$~s with a sampling frequency of $2$~kHz was simulated.
Additive white noise $x_n(t)$, with signal to noise ratio $5$, was mixed to the simulated signal to eliminate possible $0/0$ type divisions when evaluating the \eqref{eq:bicoherenceDef} bicoherence. Finally, broadband perturbations (bursts) were mixed to the signal. The bursts were created as a sum of $K$ Gaussian envelope functions multiplying independent $x'_n(t)$ white noise signals, which can be written in the following form:
\begin{equation}
x_p(t) = x'_n(t)\sum_{i=1}^{K}\exp{\bracket{-\dfrac{\bracket{t-t_i} }{2\sigma} }}.
\end{equation}
The $x'_n(t)$ (signal to noise ratio of the white noise multiplier) was chosen to be $0.5$, and the ($\sigma$) characteristic width of a single perturbation was chosen as $\sigma = 100$~ms.
The $x(t)$ signal for the bicoherence analysis is the sum of these three components:
\begin{equation}
x(t) = x_1(t) + x_n(t) + x_p(t).
\end{equation}

We applied Short Time Fourier Transformation \cite{mallat99wavelet} to visualize the time-frequency evolution of the generated signal and its components.
In the following subsectionsections we will demonstrate the effects of transients on the calculated bicoherence, and discuss the results of confidence filtering.

\subsection{Linear case with bursts}\label{sec:linear-bursts}

\begin{figure}[!hbt]
	\centering
		\includegraphics[width=0.85\linewidth]{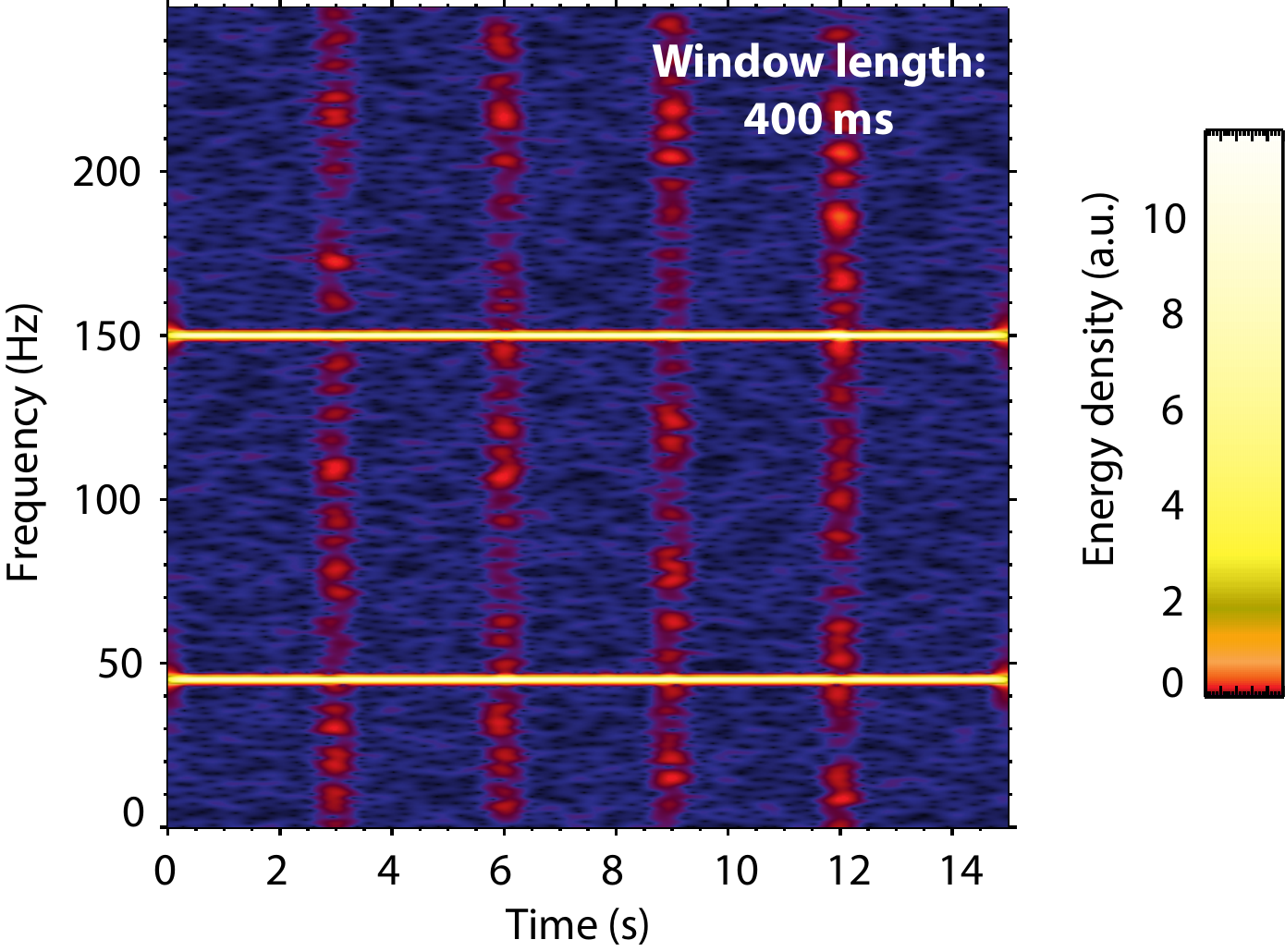}
		\caption{Spectrogram calculated for the linear case, $E=0$ with $f_1=45$~Hz and $f_2= 150$~Hz basic frequencies and 4 broadband perturbation. }
	\label{fig:linearSystemSpectro}
\end{figure}

The first natural test case is one without nonlinearities ($E=0$), and $K=4$ broadband perturbations. We will use this case to demonstrate the effects of transients on bicoherence calculation. On the spectrogram in figure~\ref{fig:linearSystemSpectro} we can observe the two eigenfrequencies at $f_1 = 45$~Hz and $f_2 = 150$~Hz, and the $4$ broadband perturbations that appear around $t\simeq 3,6,9\,\mathrm{and}\, 12$~sec. 

\begin{figure}[!hbt]
	\centering
		\includegraphics[width=0.85\linewidth]{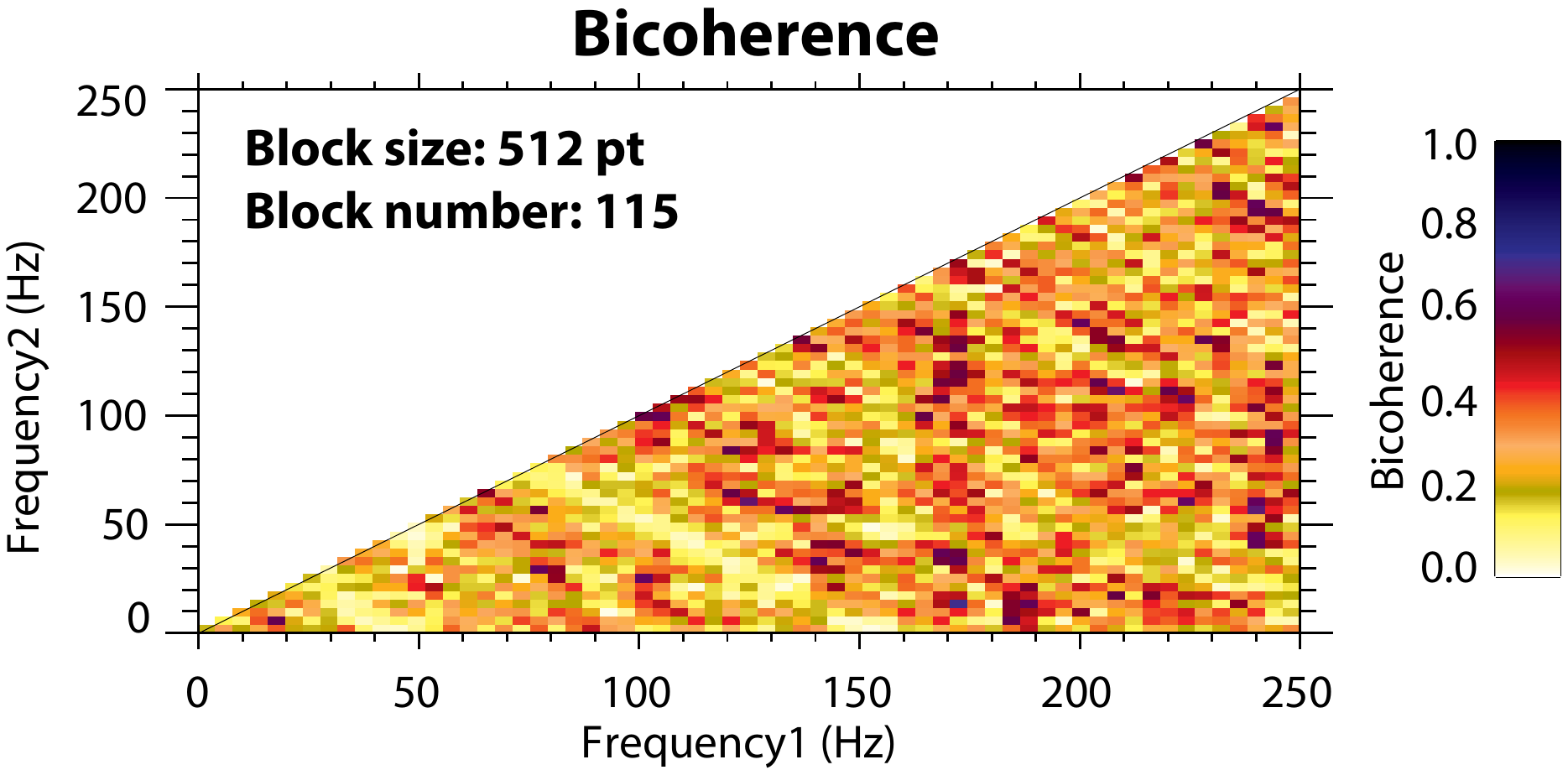}
		\caption{Lower part of the figure~\ref{fig:bicoherenceSymmetries} \textbf{P} bicoherence matrix plotted (zoom) for the linear case with broadband perturbations. High bicoherence is observed throughout the frequency plane even without phase coupling.}
	\label{fig:linearSystemBicoh}
\end{figure}

The results of bicoherence calculation, using $N=115$ blocks is shown in figure \ref{fig:linearSystemBicoh} (Only the relevant part of the $P$ matrix is plotted, indicated with the yellow dashed triangle in figure \ref{fig:bicoherenceSymmetries}).  We can observe, that even in the lack of phase-coupling there are several frequency points with high bicoherence values. Since we know that our physical system is linear, the main source of these high values are the amplitude perturbations. We now apply the procedure outlined in section~\ref{sec:nonstat-bicoh} in order to attempt the identification of false positives.

We have calculated $R = 2000$ realizations with the real signal amplitudes to generate the $\rho(b)$ distribution functions for the entire frequency plane. 
\begin{figure}[!hbt]
	\centering
		\includegraphics[width=0.85\linewidth]{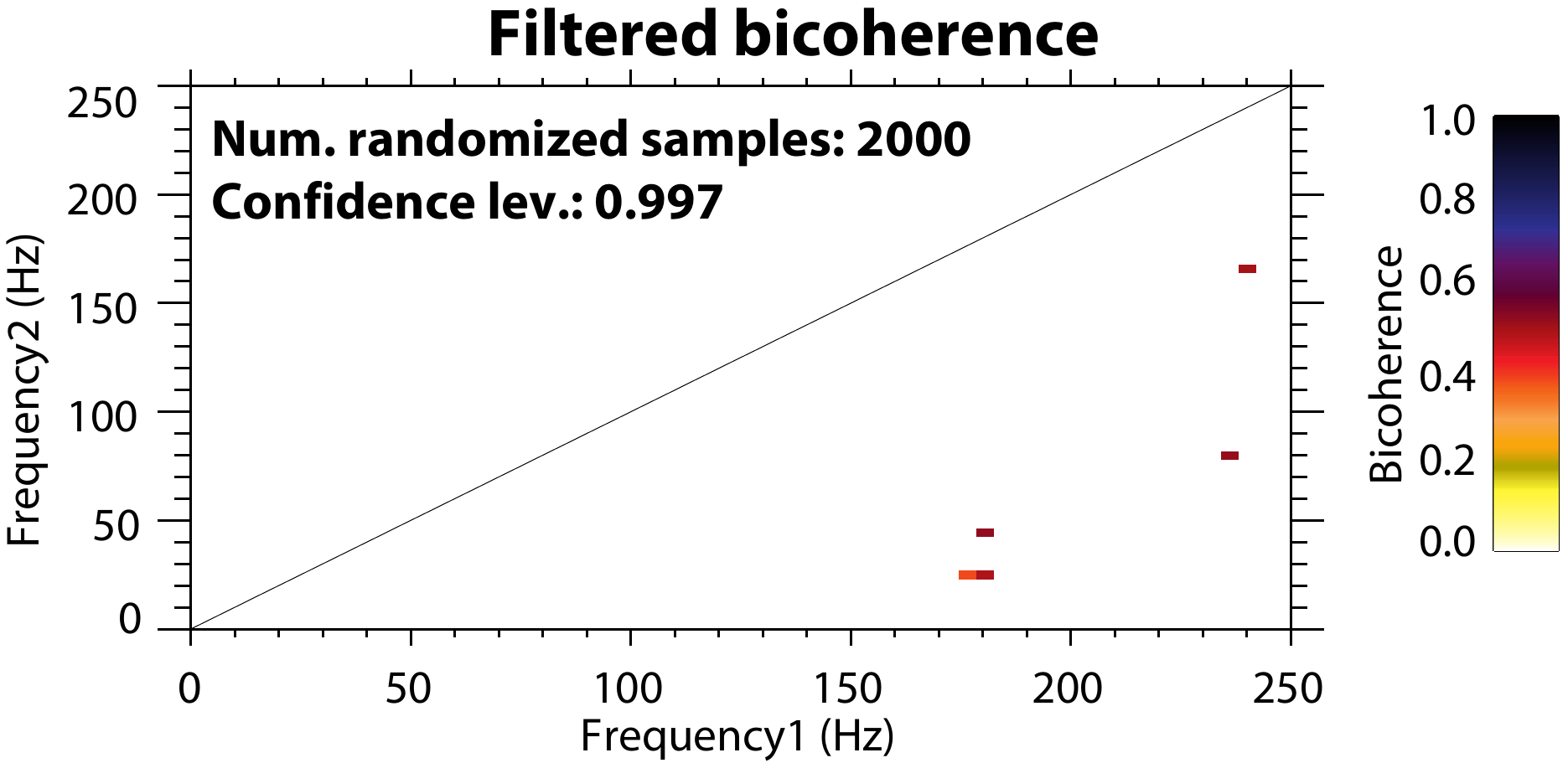}
		\caption{Linear, stationary system with $f_1=45$~Hz and $f_2=150$~Hz basic frequencies and 4 broadband perturbation. Filtered bicoherence figure shows no bicoherence at confidence level $\alpha=0.997$.}
	\label{fig:linearSystemFiltered}
\end{figure}
Using a confidence level filtering with $\alpha=0.997$, we can eliminate any $b_{f_1,f_2}^m < b_{f_1,f_2}^c(\alpha)$ points as false positives, and as shown in figure~\ref{fig:linearSystemFiltered}, most of the points disappear, as expected in the lack of phase coupling. (The selected level of $\alpha=0.997$ is considered high, but as we will show later in this section, ``real'' bicoherence remains unfiltered even at a high $\alpha$ value.)

Some unfiltered points are still left after filtering at any $\alpha$ confidence level. The origin of these is the statistical nature of the procedure: at a filtering level of $\alpha=0.997$ we will make mistakes with roughly $1-\alpha = 0.003$ probability. A bicoherence matrix for $n$ samples will contain $\mathcal{O}(n^2)$ elements, therefore we expect $n^2{\cdot}(1-\alpha)$ false positives at $\alpha = 0.997$ to be unaffected by the filtering procedure in the entire bicoherence matrix. Due to symmetries and plotting limits, in figure~\ref{fig:linearSystemFiltered}, we expect $512^2/16/8\cdot(1-0.997)\simeq 6$ false positives, which is in good agreement with the $\simeq 5$ number of points observed in the figure.
Furthermore, we have signals of finite length, which will lead to the broadening of the peaks of the Fourier spectrum. Therefore, in the case of phase coupling, we expect to have a broader, contiguous area with high bicoherence, instead of randomly spread points.

\subsection{Nonlinear case with bursts}\label{sec:nonlinear-bursts}
\begin{figure}[!hbt]
	\centering
		\includegraphics[width=0.85\linewidth]{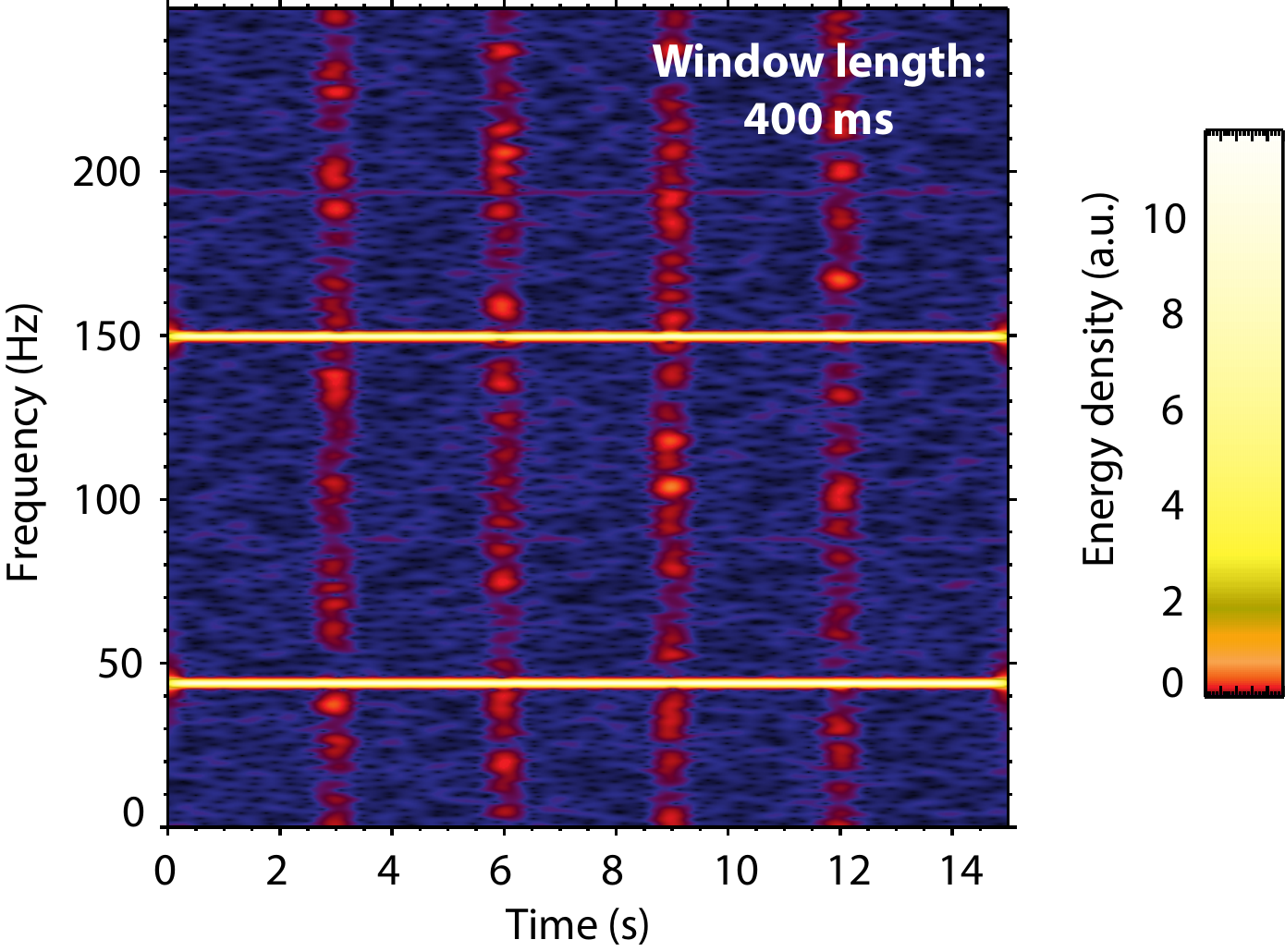}
		\caption{Spectrogram calculated for the nonlinear test case, with $f_1=45$~Hz and $f_2= 150$~Hz basic frequencies and 4 broadband perturbation.}
	\label{fig:nonlinearSystemSpectro}
\end{figure}

The second test case used the same eigenfrequencies ($f_1~=~45$~Hz, $f_2=150$~Hz) and the same 4 broadband perturbations, but with the added nonlinearity. The $E$ parameter was chosen to cause $60\%$ deviation in the force at maximum displacement compared to the linear case. The corresponding spectrogram is shown in figure~\ref{fig:nonlinearSystemSpectro}, where again we can observe the broadband perturbations, the basic harmonics and at the sum frequency of these ($f_3~=~195$~Hz) a weak frequency component appears due to the introduced nonlinear coupling.

\begin{figure}[!hbt]
	\centering
		\includegraphics[width=0.85\linewidth]{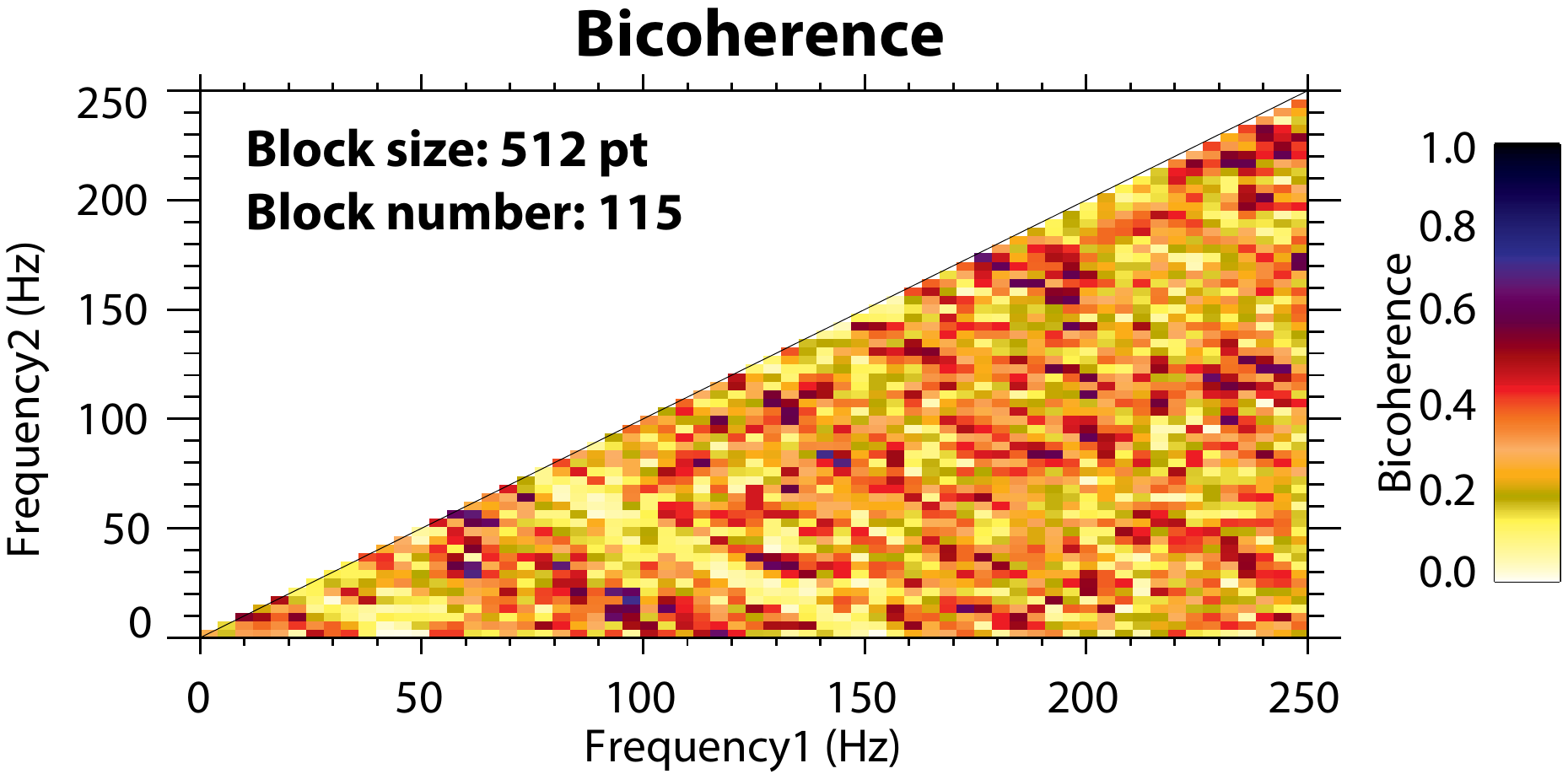}
		\caption{Bicoherence matrix calculated for the nonlinear case, including perturbations. High bicoherence appears all over the frequency-frequency plane.}
	\label{fig:nonlinearSystemBicoh}
\end{figure}

The result of bicoherence calculation is presented in figure~\ref{fig:nonlinearSystemBicoh}. Although in the model system we only define nonlinearity between the basic frequency components, high bicoherence appears at a large number of frequencies. Based on this picture alone, we cannot distinguish false high values from the ones caused by the actual nonlinear coupling.

\begin{figure}[!hbt]
	\centering
		\includegraphics[width=0.85\linewidth]{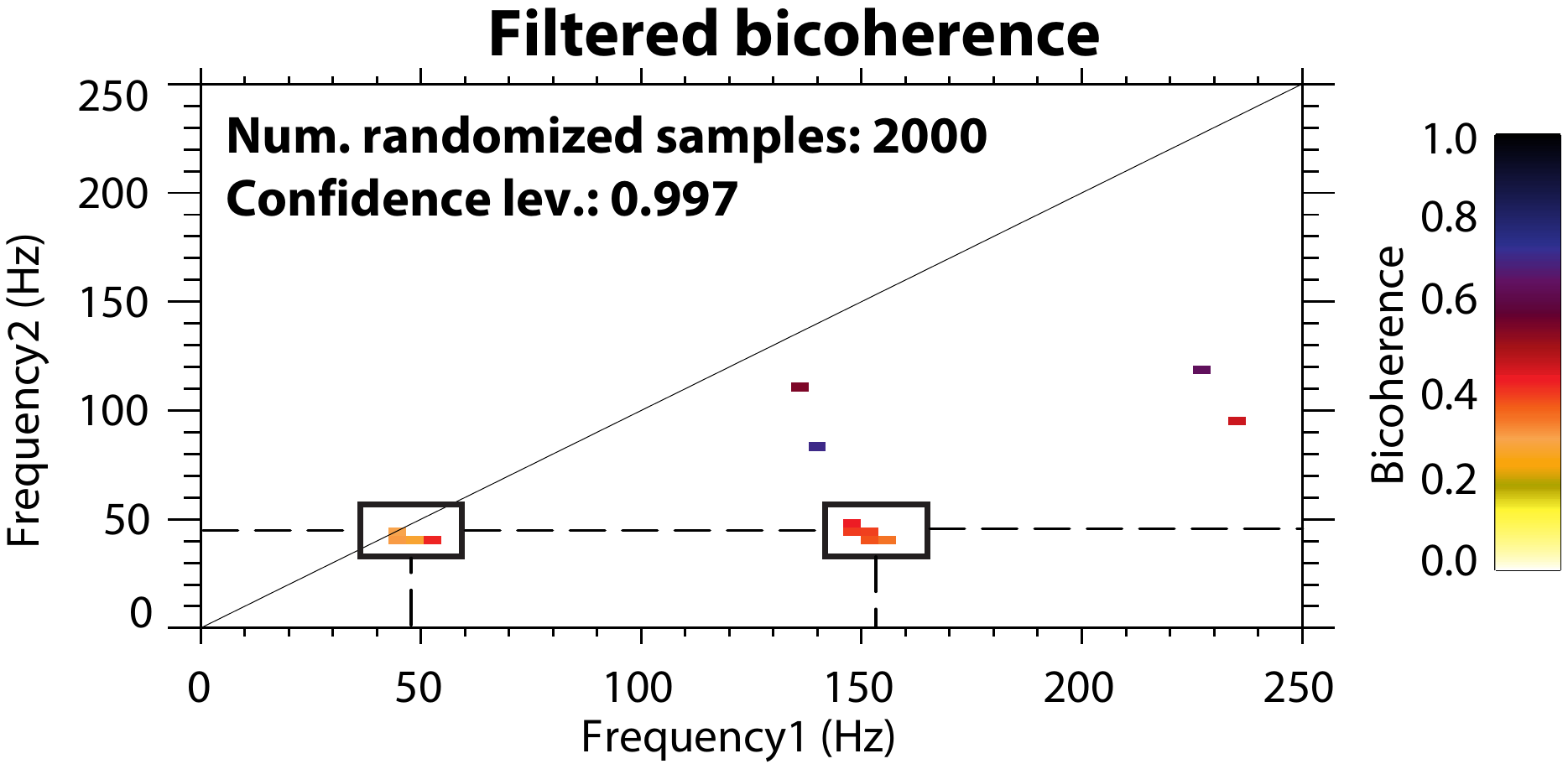}
		\caption{Nonlinear, stationary system with $f_1=45$~Hz and $f_2=150$~Hz basic frequencies and 4 broadband perturbation. Filtered bicoherence at confidence level $\alpha=0.997$ shows high bicoherence at $(45,45)$~Hz and $(150,45)$~Hz.}
	\label{fig:nonlinearSystemFiltered}
\end{figure}

The filtering process was carried out, as described above, with the result shown in figure~\ref{fig:nonlinearSystemFiltered}. Apparently, $\rho(b)$-based filtering eliminated almost all bicoherence values, except a small number of residual points (as explained earlier), and the ``true'' bicoherence corresponding to the nonlinearity of the system. The $45$~Hz frequency component is strongly nonlinear, as evidenced by high bicoherence appearing around the $(45,45)$ Hz region. We can also see the indication of significant interaction between the the $45$~Hz componentand  the $150$~Hz component, as high bicoherence appears around $(150,45)$ Hz, at a confidence level of $\alpha = 0.997$. This example illustrates that the filtering method based on the estimated $\rho(b)$ distribution can distinguish between bicoherence caused by actual phase coupling, and false positives introduced by the nonstationarity of the signal, thereby greatly aiding the interpretation of the results. 

\vspace*{-0.5cm}
\section{Discussion}\label{sec:discussion}
The evaluation method of bicoherence for nonstationary signals presented in this paper can be generalized to arbitrary n-th order coherence, and to any signal processing technique where time-averaging of complex spectra plays a role. Particularly, the coherence between two signals characterizing the linear coupling can be calculated this way by substituting the bispectrum to the cross-spectrum, and calculating the {\em random coherence probability density function}. The main idea is, that if we can formulate 'randomness' in a system in the quantity in which we are looking for systematic behaviour, then this random probability density can be compared to the actually measured quantity, and draw conclusions on its significance.

It is important to note, that the bicoherence calculation presented in the paper is ,,blind" to higher than second order nonlinearities. In case the second order nonlinearity is forbidden due to symmetry reasons, it might be necessary to study the third order nonlinearity by tricoherence \cite{chandran1994statistics}. Also one could aim for a complete description of the system by studying many levels on nonlinear terms at the same time. All of these techniques are based on detecting different higher order phase couplings, and so they are suitable for the type of significance filtering presented in the paper.

Finally, a word on the confidence level. It is straightforward to select high confidence levels in our analysis. However, one should be aware that while this helps to filter out more and more false positives, -- which corresponds to eliminating errors of first kind while testing the null hypothesis of the calculated bicoherence being the result of a random process, -- this also can cause the elimination of points exhibiting real phase coupling, -- which are errors of the second kind. Selecting the right level of confidence level requires the balancing of this trade-off, and may necessitate detailed Monte-Carlo simulations of the partially coupled systems. Choosing a moderate confidence level will cause some false positives to be sustained, which can be discriminated based on their scattered spatial distribution.

\section{Conclusion}\label{sec:conclusion}

The lowest order nonlinearity in physical systems is frequently the second-order nonlinearity, which leads to phase coupling of signal components of different frequency.
The classical bicoherence calculation was originally developed to investigate quadratic, second-order coupling in nonlinear systems, which are stationary from the time scale of the selected time windows upto the time scale of a large enough number of time windows to provide a convergent unbiased estimate.
In real world applications however, the requirement for long stationary signals for the data processing cannot always be fulfilled, and applying bicoherence calculation in such cases can lead to the appearance of false positives: high bicoherence values even in the lack of nonlinear coupling.

This paper introduces a possible way to identify false positives in the estimated bicoherence, caused by nonstationary signals. The approach is based on a Monte Carlo method, where test signals are constructed with the measured Fourier amplitudes but with random phases, and a sufficiently large ensemble of these signals is used to generate the {\em random bicoherence probability density function} ($\rho_{f_1,f_2}(b)$) for each point of the entire frequency-frequency plane. Comparing the value of bicoherence estimated from the real signal to the critical bicoherence level calculated from $\rho_{f_1,f_2}(b)$ at a given $\alpha$ confidence level for each frequency-frequency point can help decide if the estimated bicoherence is really due to second order phase coupling, or is it just an artefact caused by the changes of Fourier amplitudes in the investigated time window.

The method was tested using numerical simulations of physical test systems. We demonstrated that when the requirements for stationary signals are not met false positives emerge throughout the calculation domain. The method presented in the paper helped identify these false positives at a high confidence level (in the presented examples $\alpha = 99.7\%$) while retaining actual physical nonlinearities which were deliberately introduced in the model system for the purpose of the tests. The filtering process therefore provides an opportunity to make the difference between actual phase coupling and false positives.

\appendices
\section{Symmetries of the bispectrum}\label{sec:symmetries}
From the \eqref{eq:bispectrumDef} definition it follows that the bispectrum has several symmetries when the signal is real. 
\begin{align}
	&\underline{B(f_1,f_2)} = \Eoper{X(f_1)X(f_2)X^*(f_1+f_2)} =&  \nonumber \\
	&\Eoper{X(f_2)X(f_1)X^*(f_2+f_1)} = \underline{B(f_2,f_1)} \label{eq:symmetry1} \\[1.5ex]
	&\underline{B(f_1,f_2)} = \Eoper{X(f_1)X(f_2)X^*(f_1+f_2)} =& \nonumber \\
	&\Eoper{X^*(f_1)X^*(f_2)X(-f_1-f_2)} = \underline{B^*(-f_1,-f_2)} \label{eq:symmetry2} \\[1.5ex]
	&\underline{B(f_1,f_2)} = \Eoper{X(f_1)X(f_2)X^*(f_1+f_2)} =&  \nonumber \\
	&\Eoper{X(-f_1-f_2)X(f_2)X^*(-f_1) } = \underline{B(-f_1-f_2,f_2)} \label{eq:symmetry3} \\[1.5ex]
	&\underline{B(f_1,f_2)} = \Eoper{X(f_1)X(f_2)X^*(f_1+f_2)} =& \nonumber \\
	&\Eoper{X(f_1)X(-f_1-f_2)X^*(-f_2)} = \underline{B(f_1,-f_1-f_2)} \label{eq:symmetry4}
\end{align}
These symmetries reduce the range in which the bispectrum (or bicoherence) has to be plotted in. Let us now refer to figure~\ref{fig:bicoherenceSymmetries}. \eqref{eq:symmetry1} expresses the symmetry vs the mirroring across the $f_2 = f_1$ line. \eqref{eq:symmetry2} expresses the symmetry vs the mirroring across the $f_2 = - f_1$ line. \eqref{eq:symmetry2} \& \eqref{eq:symmetry3} leads to the equivalence of regions {\bf P} and {\bf Q} in figure~\ref{fig:bicoherenceSymmetries}, while the equivalence of regions {\bf P} and {\bf R} can be shown by applying \eqref{eq:symmetry1}, \eqref{eq:symmetry2} \& \eqref{eq:symmetry4}.

\section*{Acknowledgment}
This work has been carried out within the framework of the EUROfusion Consortium and has received funding from the Euratom research and training programme 2014-2018 under grant agreement No 633053. The views and opinions expressed herein do not necessarily reflect those of the European Commission.

\ifCLASSOPTIONcaptionsoff
  \newpage
\fi

\end{document}